
%
\input phyzzx87    


\def\R{\hbox{\rm I\kern-2pt R}}
\def\Z{\hbox{\rm Z\kern-3pt Z}}

\def\tf{\psi_{\alpha\beta}}	
\def\a{\alpha}			
\def\b{\beta}			
\def\g{\gamma}			
\def\r{\rho}			
\def\A{\cal A}			
\def\M{\cal M}			
\def\IM{{\cal I}_M}		
\def\IP{{\cal I}_P}		
\xdef\l{{(\ell)}}		

\def\refmark#1{{[#1]}}				
\REF\SEEWITa{E.~Witten, Nucl.~Phys.~B311 (1988/89) 46}
\REF\SEEWITb{E.~Witten, Nucl.~Phys.~B323 (1989) 113}
\REF\SEEMES{G.~Mess, Institut des Hautes Etudes Scientifiques preprint
                                                    IHES/M/90/28 (1990)}
\REF\SEESOR{R.~D.~Sorkin, Phys.~Rev.~D33 (1986) 978}
\REF\SEEHOR{G.~T.~Horowitz, Class.~Quantum Grav.~8 (1991) 587}
\REF\SEEGH{G.~W.~Gibbons and J.~B.~Hartle, Phys.~Rev.~D42 (1990) 2458}
\REF\SEEFUJ{Y.~Fujiwara, S.~Higuchi, A.~Hosoya, T.~Mishima and M.~Siino,
              Phys.~Rev.~D44 (1991) 1756 \ ; Phys.~Rev.~D44 (1991) 1763}
\REF\SEECARb{S.~Carlip, Class.~Quantum Grav.~8 (1991) 5}
\REF\SEEGOLb{W.~M.~Goldman, Geometric structure on manifolds and varieties
             of representations, {\it in} Geometry of group representation,
             Contemporary Mathematics vol.74,
             eds. W.~M.~Goldman and A.~R.~Magid
             (American Mathematical Society, Providence, Rhode Island, 1988)}
\REF\SEEHIR{F.~Hirzebruch, Topological methods in algebraic geometry
                                            (Springer-Verlag, Berlin, 1966)}
\REF\SEEWOO{J.~W.~Wood, Comm.~Math.~Helv.~51 (1976) 183}
\REF\SEECARa{S.~Carlip, Phys.~Rev.~D42 (1990) 2647}
\REF\SEEGOLa{W.~M.~Goldman, Invent.~Math.~93 (1988) 557}
\REF\SEECARc{S.~Carlip, University of California, Davis preprint
		UCD-91-16 (1991)}
\pubnum={178}
\titlepage
\title{Topology Change in ISO(2,1) Chern-Simons Gravity}
\author{ Kaoru Amano
 \quad and \quad Saburo Higuchi
\footnote{\star}{shiguchi@cc.titech.ac.jp}
}
\address{Department of Physics\nextline Tokyo Institute of Technology
\nextline Meguro, Tokyo,152  Japan}
\vfil
\abstract
In $2+1$ dimensional gravity,
a dreibein and the compatible spin connection can represent a
space-time containing a closed spacelike surface $\Sigma$
only if the associated SO(2,1) bundle restricted to $\Sigma$
has the same non-triviality (Euler class) as
that of the tangent bundle of $\Sigma.$
We impose this bundle condition on each external
state of Witten's topology-changing amplitude.
The amplitude is non-vanishing only if
the combination of the space topologies  satisfies
a certain selection rule.
We construct a family of transition paths which reproduce all the
allowed combinations of genus $g\ge 2$ spaces.
\endpage

\chapter{Introduction}

2+1 dimensional pure gravity as reformulated by Witten\refmark\SEEWITa \
may provide us with a tractable model of Einstein
gravity with topology change\refmark\SEEWITb.
The theory (without a  cosmological constant) can be identified with
a Chern-Simons gauge theory (CSGT) of
structure group ISO(2,1)
 (the Poincar\'e group in 2+1 dimensions)
with an appropriate Killing form.
The path integral in this ISO(2,1) CSGT is computable for an
arbitrary space-time topology with or without boundary.
Accordingly not only propagators but vacuum amplitudes, Hartle-Hawking
wave functions, and other amplitudes involving sectional topology changes
 can be constructed.  They all can be brought into the form of
the sum (integral) over the moduli space of flat connections up
to gauge transformations.
Whether an amplitude with a particular mode of topology change
vanishes or not
can be judged by seeing whether
there exists a  flat connection
that satisfies the corresponding boundary condition.

    However, Witten's topology-changing amplitude may not entirely
correspond to
the change of {\it spatial}
topology.
There is no guarantee that
the asymptotic space-time admits a spacelike slice.
This implies that the in- or out-states may not be completely free from
the interacting region, and then
it is quite awkward to regard that amplitude
as referring to observable spatial topologies.
We propose to restrict the theory so that asymptotic states are
only those with a spacelike slice.
But how can this be achieved?

   In his first paper on Chern-Simons gravity\refmark\SEEWITa,
 Witten selects
 a special sector of the ISO(2,1) CSGT
 for the description of gravity.
Based on a relation between flat SO(2,1) structure and
conformal structure on closed surfaces, he adopts, in effect, the
following restriction.
In  canonical theory on a closed 2-space $\Sigma$ of genus $g\ge2,$
of all the flat SO(2,1) connections on $\Sigma$
only those defined in an SO(2,1)
bundle with Euler class $\pm (2g-2)$ are relevant to gravity.
This condition is in fact equivalent to demanding that $\Sigma$ can be
embedded in the space-time as a spacelike hypersurface\refmark\SEEMES.

What we do is as follows.
We require that the path integral should be taken only over
connections in an ISO(2,1) bundle
whose portion over each component of space-time boundary has the right
Euler class.
Then we ask what kind of topology change is possible.
We will see that the condition for
the existence of a bundle satisfying the above
condition leads to a selection rule on the combination of
2-space topologies involved in the process.
Then we construct a family of transition paths that cover
a substantial part of allowed topology combinations.

As we do not actually work out the amplitude or check  the
consistency of the whole model, our investigation remains
at a preliminary stage.
Our work is comparable to refs.[\SEESOR][\SEEHOR][\SEEGH][\SEEFUJ].
In particular, our selection rule has a striking formal resemblance
to that of Sorkin\refmark\SEESOR, which arises from Lorentzian cobordism.
Also the pull-back construction of Horowitz\refmark\SEEHOR \
guarantees the existence
of a certain type of topology-changing paths in our approach.

Besides the issue of spatial topology change, this paper contains
a contribution towards a geometric understanding of
Chern-Simons gravity.
There are two apparently different ways to associate
an ISO(2,1) flat connection with a space-time geometry.
One is  to identify the connection with
a dreibein-spin connection pair,
and the other is to identify the holonomy of the connection with that
of a flat Lorentzian structure.
The equivalence with gravity in terms of the action and
equations of motion
thereof is based on the first,
while  the second is invoked in the original derivation of
the restriction on the SO(2,1) bundle\refmark\SEEWITa.
Whether these two principles of identification are compatible
is a non-trivial question\refmark\SEECARb.
In the following section, we demonstrate the compatibility by
deriving the second from the first.
This strengthens the interpretation of the bundle condition as
a condition for a spacelike space, which we base on the first
principle.

This paper is organized as follows.
In sect.2 we discuss how an ISO(2,1) gauge field dictates
space-time geometry, and  then clarify the condition
for space-time to contain a spacelike slice.
In sect.3 we derive the selection rule and then produce the examples of
topology-changing paths.
In sect.4 we discuss  our results.

\noindent {\it Convention} \break\hfil
Throughout this work, by SO(2,1) and ISO(2,1) we actually mean
the maximal connected subgroups instead of the full groups.
Thus SO(2,1) in our convention is isomorphic to
PSL(2,\R)=SL(2,\R)/$\{\pm 1\}$.

\chapter{ISO(2,1) connections, space-time, and spacelike surfaces}

	We shall discuss how ISO(2,1) flat connections relate
to space-time geometry.
	We will see that the the holonomy of a flat Lorentz structure
arising from a flat connection is identical
with that of the connection\refmark\SEEWITa\refmark\SEECARb.
	Also the condition for space-time to admit a closed 2-space is
introduced\refmark\SEEMES.

\section{The Einstein-Hilbert-Chern-Simons action}

	Let $A$ be a connection in an ISO(2,1) flat bundle $\hat{P}$
over an orientable 3-manifold $M$.
	Locally $A$ is a Lie-algebra-valued 1-form
$$
A = e^a P_a + \omega^a J_a,					\eqn\CON
$$
where $P_a, J_a (a = 0,1,2) $ are a basis for the ISO(2,1) algebra
with the commutators
$[P_a, P_b] =0, [J_a, P_b] = \epsilon_{abc}P^c,$ and
$[J_a, J_b] = \epsilon_{abc}J^c .$
	The indices are raised and lowered with the Lorentz metric
$(\eta^{ab}) = (\eta_{ab}) = {\rm diag} [-1,1,1].$
	$A$ is expressed as eq.\CON \  with respect to local sections
of $\hat{P}$ over coordinate patches $U_{\a}$ of $M$.
	The local sections are related to each other by transition functions
$\tf$ on overlaps $U_{\a} \cap U_{\b}$.
	Correspondingly the local expressions of $A$ satisfy the relation
$$
A_{\b} = \tf^{-1} A_{\a} \tf + \tf^{-1} d \tf.		\eqn\TRF
$$
	We can choose local sections so that all $\tf$ take values
in the SO(2,1) subgroup generated by $J_a$.
	This reduces $\hat{P}$ to an SO(2,1) bundle $P$.
	$P$ is also flat by the semi-direct product nature of the
group ISO(2,1).
	The decomposition \CON \  of $A$ into $\omega$ and $e$ parts
is now global,
with $\omega$ an SO(2,1) connection in the reduced bundle $P$,
and $e$ an associated 1-form dreibein, which may be degenerate.

	To make contact with gravity, we treat $e$ and $\omega$ as variables
for $2 + 1 $ dimensional Einstein theory in the first-order formalism.
	Then the Einstein-Hilbert action,
$$
S=\int\nolimits_M e^a (d\omega_a +{1\over 2} \epsilon_{abc} \omega^b \omega^c),
								\eqn\CSA
$$
is an ISO(2,1) Chern-Simons action\refmark\SEEWITa.
	$S$ is well-defined in our system of local sections
and is invariant under any gauge transformations
that reduce to SO(2,1) transformations on the boundary of $M$.
	By a gauge transformation, we mean the one globally defined:
a set of ISO(2,1) valued functions $h_\a$ with the consistency condition
$h_\a \tf = \tf h_\b $ acting on $A$ as
$$
A_\a \rightarrow A_\a' = h_\a^{-1} A_\a h_\a + h_\a^{-1} d h_\a.  \eqn\GTA
$$
	Chern-Simons action \CSA \  gives equations of motion
which demand that $A$ be flat,
$$
d A + A A = 0.							\eqn\FLAT
$$
	The existence of solutions is guaranteed by the flatness of $\hat{P}$.

\section{Development and holonomy}

	The geometrical consequence of eq.\FLAT \  is that
space-time $M$ with $A$ is locally Minkowski.
	That is, it can be mapped into Minkowski space $X$ preserving
its local metric content.
	We shall construct such a map.

	To begin with, we take a frame $u=(q, f_0, f_1, f_2 )$ in $X$,
where the tangents $f_a$ to $X$ at $q \in X$
are orthonormal, $f_a \cdot f_b = \eta_{ab}$.
	We let the group ISO(2,1) act on $u$ on the right by the rule,
$$
( q, f_a) \cdot T L = ( q + T^a f_a, f_b L^b{}_a) ,		\eqn\ACU
$$
with $ T = e^{T^a P_a} , \ L J_a L^{-1} = J_b L^b{}_a$.
	Now the following first-order differential equation makes sense:
$$
d u = u \cdot A.						\eqn\DIF
$$
	The flatness \FLAT \  says \DIF \  is locally integrable.
	Given a frame $u_0$ in $X$
and a local section over point $* \in M$,
we get a frame field $u(p)$ on a neighbourhood $U$ of $*$ by integrating
eq.\DIF \  along paths in $U$ starting from $*$
with the initial value $u(*)=u_0$.
In other words, by associating $p=*$ with $u=u_0$ and
moving them together according to the transportation law \DIF,
we can map $U\subset M$ onto a set of frames of $X$.
	Then the map of $U$ into $X: p \mapsto u(p) \mapsto q(p)$
preserves the metric.
	This is evident if we write down eq.\DIF \
in terms of $e$ and $\omega$:
$$
d q = e^a f_a , \; \; \; d f_a = -\epsilon_{abc} \omega^b f^c.	\eqn\DIFS
$$
	Thus we obtained a local version of the map we have been looking for.

	To get a global version, we extend the above construction to general
paths from $*$, not necessarily confined to its neighbourhood.
	In the integration of eq.\DIF, we switch local sections over the path
 if necessary by the rule consistent with \TRF:
$ u_\b = u_\a \cdot \tf {\rm \ on \ } U_\a \cap U_\b $.
	At the switching, the frame $u$ just pivots, with $q$ maintaining its
position and course, as our $\tf$ are all SO(2,1)-valued.
	To disentangle path dependence we lift the path to a universal cover
${\tilde M}$ of $M$.
	The end positions of $q$ put together give
the {\sl development map}\refmark\SEEGOLb \
$ \phi: {\tilde M} \rightarrow X $ .

	The holonomy of $A$ arises naturally in the integration of eq.\DIF \
in the form,
$u(\g \cdot * ) = u_0 \cdot \rho(\g) {\rm \ for \ } \g \in \pi_1(M,*)$.
	Let us denote by the same $\rho(\g)$
also the Poincar\'{e} transformation on
$X$ whose differential map sends $u_0$ to $u_0 \cdot \rho(\g)$.
	Then for any $\g \in \pi_1(M,*)$ and ${\tilde p} \in {\tilde M}$,
we have,
$$
 \phi(\g \cdot {\tilde p}) = \rho(\g) \phi({\tilde p})    .	\eqn\HOL
$$
	Thus the holonomy of connection $A$ is also
that of space-time geometry.
	The gauge transformation $h=\{h_\a\}$ acts on $u$ as
$ u_\a \rightarrow u'_\a = u_\a \cdot h_\a$ for each local section.
	This is intuitively a nonuniform Poincar\'{e} transformation.
	It deforms the development map $\phi$ rather arbitrarily except that
the holonomy homomorphism
$\r: \pi_1(M,*) \rightarrow {\rm ISO(2,1)}$
changes only by an overall conjugation,
$\rho \rightarrow \rho' = h_0 \rho h_0^{-1}$
with some $h_0 \in {\rm ISO(2,1)}$.

\section{The space condition}

	Let $\Sigma$ be a closed orientable surface (of genus $g$) embedded
in $M$.
	Suppose $\Sigma$ is spacelike in a space-time associated with $A$,
a flat connection in $\hat P$.
	That is, there exists a gauge choice such that the dreibein $e$
induces a positive definite metric on $\Sigma$.
	Then as we shall see (sect.2.3.2), the Euler class (sect.2.3.1)
of $P|_\Sigma$ the restriction to $\Sigma$ of the reduced SO(2,1)
bundle $P$ (sect.2.1), is given by\refmark\SEEMES,
$$
 {\rm eul}(P |_\Sigma) = \pm \chi_\Sigma,		\eqn\SLC
$$
where $\chi_\Sigma = 2 - 2g $ is the Euler characteristic of $\Sigma$.
	In this statement $P$ can be replaced with the original ISO(2,1)
bundle $\hat P$  without changing the content.

\subsection{2.3.1.~Euler class}
	Let $G$ be one of ISO(2,1), SO(2,1), GL${}^+$(2,\R), SO(2),
or any other
groups which have the same homotopy type as the circle $S^1$,
and let $E$ be a
$G$ bundle over $\Sigma$.
	The Euler class ${\rm eul}(E) \in \Z \approx {\rm H}^2(\Sigma,\Z)$
classifies $E$ completely (up to isomorphism).
	It gives a one-to-one correspondence between integers and the bundles
with fixed $G$ and $\Sigma$.
	Another qualification is that ${\rm eul}(E)$ measures the obstruction
to taking a global section of $E$.
	To give a more definite idea, we remove a disc from $\Sigma$
and cover the hole by a larger open disc $D$, so that
$\Sigma = C \cup D$
where $C$ is the compliment of the smaller disc.
	We take sections over the patches $C$ and $D$, obtaining transition
functions $ \psi_{CD} $ and $\psi_{DC} = \psi_{CD}{}^{-1}$ on $C\cap D$.
	The integer ${\rm eul}(E)$ is given by counting how many times
$\psi_{CD}$ winds in $G \simeq S^1$ as we go round the annulus $C \cap D$.

	A vanishing ${\rm eul}(E)$ will mean that the $E$ admits a global
section, i.e., it is trivial.
	In other cases the sign of {\rm eul} depends on the orientation of
$\Sigma$.
	Also the following should be obvious.
	If $G$ bundle $E$ reduces to an SO(2) bundle $E_0$,
i.e., $E$ and $E_0$ share a common system of transition functions, then
${\rm eul}(E) = {\rm eul}(E_0)$.
	Structure group $G$ is reducible to its SO(2) subgroup.
	So there is an $E_0$ for every $E$.
	This allows us to rephrase the above procedure into a Gauss-Bonnet type
formula: Picking an arbitrary connection $\nu$ in $E_0$, we have,
$$
{\rm eul}(E) = {1 \over 2 \pi i} \int\nolimits_\Sigma d\nu,	\eqn\GBT
$$
where we used the U(1) notation, $\nu = i \nu_{12}$.

	We will need the following fact shortly:
${\rm eul}(\theta_\Sigma) = \chi_\Sigma$\refmark\SEEHIR.
	This is the special case of eq.\GBT \ when $E = \theta_\Sigma$ is the
GL${}^+$(2,\R) bundle defined by the transition functions
$ \tf = (\partial \phi^i_\a / \partial \phi^j_\b)$,
the Jacobian matrices for a consistently oriented system of coordinates
$\{ \phi_\a \}$ of $\Sigma$.
	$\theta_\Sigma$ is associated with the tangent bundle
$T\Sigma$, and
is itself called the tangent bundle of $\Sigma$.

\subsection{2.3.2.~Proof of eq.\SLC}
	The assertion is that $P | _\Sigma$ and the tangent bundle
$\theta_\Sigma$ have the same Euler class up to a sign.

	We may think of ${\tilde \Sigma}$ as embedded in Minkowski space $X$,
spacelike and with SO(2,1) local frame fields $f_a$ on it, in accordance
with the construction of sect.2.2.
	The local sections of $P |_\Sigma$
can be adjusted so that $f_0$ is normal to ${\tilde \Sigma}$ everywhere.
	Then $f_0$ is common to all the sections over overlapping patches,
and $f_1$ and $f_2$ constitute an SO(2) frame tangent to ${\tilde \Sigma}$.
	The transition functions for $P |_\Sigma$ are now
also those for the tangent bundle $\theta_\Sigma$ of $\Sigma$ or of the same
surface with a reversed orientation.
	Thus these two bundles reduce to a common SO(2) bundle, and therefore
have a common Euler class.
	So
${\rm eul}(P |_\Sigma) = \pm {\rm eul}(\theta_\Sigma)
				    = \pm \chi_\Sigma  $ .
We are done.

\subsection{2.3.3.~Comments}
	In the above we used the notation from the construction of a
development map, but it is not essential.
	With a little modification the proof applies to the bundle of
SO(2,1) frames tangent to a not necessarily flat Lorentzian 3-manifold with
a spacelike hypersurface $\Sigma$.
	Thus eq.\SLC \ holds in a more general situation than suggested
before.

	Sticking with our original motivation, we require $P$ to
be flat.
	$P|_\Sigma$ is also flat then.
	For $g \geq 2$, flat bundles $E$ over $\Sigma$ with structure
group SO(2,1)
are those with
$ \chi_\Sigma \leq {\rm eul}(E) \leq - \chi_\Sigma$
\refmark\SEEWOO\refmark\SEEMES.
$P|_\Sigma$ must belong to this range
and the space condition \SLC \ picks out the ones with the maximal absolute
value.
	For $g = 0,1$, flat $E$ is necessarily trivial, ${\rm eul}(E) = 0$.
	So for $g=0$, no $P$ is allowed.
	This reflects the obvious fact that sphere ${\tilde \Sigma} = \Sigma$
cannot be embedded in $X$ spacelike.

	The condition \SLC \ is necessary for $\Sigma$ to become spacelike.
	A natural question is then, is it sufficient?
	To give a partial answer, we restrict ourselves to the topology
$M = \Sigma \times \R$.
	This is a suitable space-time topology for canonical formulation,
taking $\Sigma $ at some time as an initial surface.
	$P$ is determined by its restriction to the initial surface.
	The condition \SLC \ specifies $P$.
	The holonomy homomorphism
$\r: \pi_1(M) = \pi_1(\Sigma) \rightarrow {\rm ISO(2,1)}$
should contain all gauge invariant information on the flat connection.
(This view entails the assumption that gauge transformation at infinity
$t\rightarrow \pm\infty$ is subject to no restrictions.)
	Then we may say the condition \SLC \ is sufficient if
for any flat connection in the selected $P$ there is an
appropriate flat Lorentzian manifold that admits spacelike $\Sigma$ and
has the holonomy corresponding to the connection.
	For it implies that in some gauges the connection reproduces that
Lorentzian manifold with $\Sigma$ embedded in the desired way.
	Mess\refmark\SEEMES \ classifies flat Lorentzian manifolds
containing a spacelike hypersurface by their holonomy.
	He establishes that for $g \geq 2$, the moduli space of flat
connections in the ${\rm eul} = \pm \chi_\Sigma$ bundles parametrizes
a certain family of reasonable space-times with a spacelike slice.
	Thus the space condition \SLC \ is sufficient in the above sense
for $g \geq 2$.
	For $g=1$, the only allowed $P$ is a trivial bundle.
	In this bundle some flat connections correspond to space-times with
spacelike $\Sigma$ but others do not.
	For the necessary restriction on the space of flat connections,
see refs.\refmark\SEEMES\refmark\SEECARa.

\chapter{Quantization, transition paths, and a selection rule}
	We regard gravity as dynamics of space geometry,
and restrict the species of spaces to be closed orientable surfaces.
	The space-time metric must be such that the spaces are spacelike
by the metric induced on them.
	In the present theory the classical space-time to accommodate
space dynamics seems to invariably have a constant spatial topology.
	The space-time is flat Lorentzian, and time-orientable because
our structure group is a connected subgroup
of the Poincar\'e group.
	Mess\refmark\SEEMES \
proves that a compact, flat, time-orientable Lorentzian manifold
with spacelike boundary necessarily
has a topology $M={\Sigma}\times [0,\, 1]$, with
$\Sigma$ spacelike at each time $t \in [0,\, 1].$
	We interpret this as demonstrating that topology change is not
allowed in the ISO(2,1) gravity at classical level.
	To put it the other way round, our classical topology change is
what is ruled out by Mess' theorem: either the space-time has a topology
different from ${\Sigma}\times [0,\, 1]$
or defies a spacelike time-slicing.
	Note that we are assuming the non-degeneracy of space-time metric.
	Without this assumption, there do exist solutions
in the ISO(2,1) CSGT that can almost be called
space-times with spatial topology change\refmark\SEEHOR.
	Nonetheless those solutions are presumably more suitable
as intermediate paths than
as full-fledged classical space-times.
	We prefer to
consider such solutions in the context of quantum theory.

\section{Quantization\refmark\SEEWITa\refmark\SEEWITb}

	The ISO(2,1) CSGT owes much of its simplicity to the fact that the
structure group
$\hat{G} = {\rm ISO(2,1)}$
is equal to the total space of the tangent bundle to
$G = {\rm SO(2,1)}$
as a Lie group:
$\hat{G} = T G$.
	This leads to a relation of the form
$\hat{\A} = T \A$,
where $\hat{\A}$ is any of the spaces of connections, flat connections
in the ISO(2,1) bundle $\hat{P}$, or the corresponding moduli spaces,
and $\cal A$ the counterpart for the reduced SO(2,1) bundle $P$.
	(Of course we cannot expect the relations to be precise when
$\cal A$ is not a manifold.)

	In canonical formulation on the surface $\Sigma$, the relation
$ \hat{\M} = T \M$
is particularly important,
where $\hat{\M}$ is the moduli space of flat connections in $\hat{P}$
and $\M$ counterpart for $P$.
	The $\hat{P}$ and $P$ here can be thought of
as either bundles over the space-time manifold
$ M = \Sigma \times \R $,
or those over the 2-space $\Sigma$.
	$\hat{\M}$, the moduli space of classical solutions, is the
physical phase space.
	The symplectic structure from the action \CSA \ says
$e$ and $\omega$ are canonically conjugate.
	The relation
$ \hat{\M} = T \M$
suggests we should take a gauge equivalence class
${\rm cls\ }\omega \in \M $ as `coordinates'
and tangents to $\M$ as `momenta.'
	Correspondingly a physical state in canonical quantization is
represented by a function on $\M$, or equivalently
by a gauge-invariant function of connection $\omega$ in $P$.
	For $g = 1$, some modification is necessary since the space
of flat connections must be restricted\refmark\SEECARa.
	For $g \geq 2$, with $P$ satisfying the space condition \SLC \ ,
$\M$ can be identified with Teichm\"{u}ller space of $\Sigma$.
	This arises from the fact that the holonomy homomorphism
of a flat connection in $P$ gives a discrete embedding of
$\pi_1(\Sigma)$ into SO(2,1) $\approx$ PSL(2,$\R$),
and vice versa\refmark\SEEGOLa.
	In particular, to provide $\Sigma$ with the structure
of a Riemann surface is effectively to give a flat connection in $P$
up to gauge transformation.
	We will need these facts later.

	To be precise the above procedure does not complete canonical
quantization.
	For we would have to discuss observables and the measure on $\M$,
which includes taking care of the mapping class group
on $\Sigma$\refmark\SEECARa\refmark\SEECARb\refmark\SEECARc.
	We will not go into these issues.

	Now we proceed to the path integral approach.
	We take space-time 3-manifold $M$ to be a compact orientable manifold
with boundary consisting of connected components $\Sigma_1, \ldots \Sigma_N$,
each of which is closed and orientable.
	With some SO(2,1) flat connections $\omega_j$ given on $\Sigma_j$,
we consider the path integral,
$$
 \IM (\omega_1, \ldots, \omega_N)
        = \int\nolimits_{\omega =\omega_j {\ \rm on\ } \Sigma_j} D\omega \int
        D e \; \exp iS.
								\eqn\TRA
$$
	The first integral is over all SO(2,1) connections $\omega$ on $M$
that gives $\omega_j$ when restricted to $\Sigma_j$,
and the second over all dreibeins $e$ associated with the same bundle
as $\omega$ belong to.
	The integral region has a tangent bundle structure with its base
restricted by the boundary conditions.
	The integral on $e$ over fibres can easily be done to give,
$$
 \IM (\omega_1, \ldots, \omega_N)
        = \int\nolimits_{\omega =\omega_j {\ \rm on\ } \Sigma_j}
                     D\omega\      \delta [d \omega + \omega \omega],
								\eqn\INW
$$
where the delta functional has its support on flat $\omega$.
	Let $P_j$ denote the SO(2,1) bundle over $\Sigma_j$ to which $\omega_j$
belongs.
	The integral \INW \ vanishes unless $\omega_j$ have some flat $\omega$
as a common extension.
	Let us assume the existence of such an extension.
	Let $P$ be the flat SO(2,1) bundle over $M$ for the extension,
and $\IP$ the restriction of the integrand \INW \ to connections in $P$.
	If there is more than one $P$ then $\IM$
is the sum of $\IP$.
	$\IP$ has an unphysical divergence arising from gauge invariance,
which is removed by an appropriate gauge fixing.
	Another possible source of divergence is variations
of flat $\omega$ that keep $\omega$ flat but are transverse to gauge orbits.
	In other words, $\IP$ diverges if the moduli space of flat connections
in $P$ has at least one dimension.
	Geometrically this will occur if the interior of $M$
has a homotopically non-trivial loop `of its own' in the sense that
the holonomy around it cannot be controlled by the holonomy on the boundary.
	Physically it is an infrared divergence due to a portion of space-time
escaping from the Planckian into the classical dimensions\refmark\SEEWITb.

	We would like $\IM$ to represent the amplitude of the process
involving 2-spaces $\Sigma_j$ mediated by the 3-manifold $M$.
	For this we would like the state attached
to each boundary component $\Sigma_j$ to admit an interpretation
as a space-time with a spacelike slice.
	We therefore require $P_j$ to have Euler class $\pm \chi_j$
where $\chi_j = \chi_{\Sigma_j}$.
	This effectively demands the space condition \SLC \ for the $P$
over $M$ that accompanies the extension of $\omega_j$.
	This does not mean however, that we seek to interpret the whole $M$
as a Lorentzian manifold with spacelike boundary.
	This is usually impossible as noted earlier.
	Our basic view is that the space-time geometry of asymptotic states
is observable but not is that of intermediate ones.

	For convenience we will say a state on a surface $\Sigma$ is
in the {\it spatial sector}, or just {\it spatial},
if the corresponding $P|_\Sigma$ satisfies the
requirement ${\rm eul}(P|_\Sigma) = \pm \chi_{\Sigma}$.
Similarly an amplitude is said to be in the spatial sector when all
the external states are spatial.

\section{A selection rule}

	If there is no restriction on the external states,
then for any $M$, ${\cal I}_M$ is not zero for some states.
	This is because there exists
at least one SO(2,1) flat bundle over $M$,
the trivial bundle for instance,  hence a flat connection $\omega$.
	The amplitude does not vanish for $\omega_j=\omega|_{\Sigma_j}.$
	Thus topology change is arbitrary.
	When restricted to the spatial sector however, the situation is
quite different.

	Consider $M$ with a topology obtained by removing from a $g = 2 $
handlebody (solid double-torus) two $g=1$ handlebodies (solid tori)
so that  the two handles are hollow (fig.1).
	The boundary of $M$ consists of $g=1$ surfaces (tori)
$\Sigma_1, \Sigma_2$, and a $g=2$ surface (double-torus) $\Sigma_3$.
	$\IM$ does not have infrared divergences since the holonomy
of flat $\omega$ on $\Sigma_3$ completely determines that on $M$.
	We now claim that
on the support of $\IM$
the state on $\Sigma_3$ cannot be in the spatial sector.
	Look at path $\g$ in $\Sigma_3$ depicted in fig.1.
	This closed path is homotopically non-trivial in $\Sigma_3$
(with the basepoint $*$) but is trivial in $M$ as it can be shrunk to $*$.
	If an $\omega_3$ on $\Sigma_3$ extends to a flat $\omega$ on $M$,
then its holonomy round $\g$ is necessarily trivial:
with its holonomy homomorphism $\rho_3$,
$\rho_3 (\g) = 1 \in {\rm SO(2,1)}$.
	This is not compatible with the space condition.
	For this condition requires $\rho_3$ to be a discrete embedding
of $\pi_1(\Sigma)$ into
${\rm SO(2,1)}\approx {\rm PSL(2,\R)}$\refmark\SEEGOLa.
	In particular, $\rho_3(\g) \not= 1$ for
$\g \not= 1 \in \pi_1(\Sigma,*)$.
	Thus we find that the $M$ is irrelevant to the processes
for the spatial
sector.
	This generalizes to any $M$ that has a boundary component of
$g\geq 2$ with a loop contractible in $M$ but non-contractible in the boundary.

	In the above we saw one way to identify processes suppressed
by the space condition, working on particular $M$.
	We now ask instead what combinations of spatial topologies can give
a non-vanishing amplitude, without specifying the interpolating manifold $M$.
	We answer this by the following.

\proclaim {Selection rule}.
Suppose the disjoint union
$\Sigma_1\sqcup\Sigma_2\sqcup\dots\sqcup\Sigma_N$
of genus $g_j\ge 1$ surfaces $\Sigma_j$ bounds some space-time
manifold $M$ such that the amplitude ${\cal I}_M$ is not
identically zero in the spatial sector.
Then the Euler characteristics $\chi_j = 2 - 2g_j$
of the closed surfaces $\Sigma_j$ satisfy the relation,
$$
 \sum\nolimits_{j=1}^N \epsilon_j \chi_j = 0, 			\eqn\SLR
$$
with some sign assignment $\epsilon_j = \pm 1$ .

	This follows from the observation that $\IM$ can be non-zero
only if the bundle $\sqcup_{j=1}^N P_j$
over the boundary $\sqcup_{j=1}^N \Sigma_j$
extends to an SO(2,1) bundle $P$ over $M$.
	In fact, take such a $P$ and reduce it to an
${\rm SO(2)} \approx {\rm U(1)}$ bundle $Q$.
	We pick an arbitrary $U(1)$ connection $\nu$
in $Q$ and apply the formula  \GBT \
over the boundary components $\Sigma_j$.
If we use the orientation of $\Sigma_j$ induced from $M$,
the sum of euler classes of $P_j$ becomes as follows:
$$
\sum\nolimits_{j=1}^N {\rm eul} (P_j)
          = {1 \over 2 \pi i} \int\nolimits_{\partial M} d \nu
          = 0.							\eqn\VEC
$$
The integral vanishes by Stokes' theorem since the curvature $d\nu$ is
a closed 2-form over $M$.
	Eq.\SLR \ follows from eq.\VEC \ with the space condition
$ {\rm eul}(P_j) = \pm \chi_j$.

	Note that the flatness of $P$ was not used in the proof.
	Actually the validity of the constraint  \SLC \
on spatial boundaries
extends to three-dimensional Lorentzian gravity in general.
	See sect.4 for a comparison with Lorentzian cobordism.

	The topology combination corresponding to the example of fig.1
gives $\chi_1 = \chi_2 = 0$, $\chi_3 = -2$.
	This is forbidden no matter what the $M$ is.
	A generalization of this situation is given by
$\Sigma_j$ of $g_j \ge 1$, with $g_N =g_1 +g_2 +\cdots + g_{N-1},\; N \ge 3$.
	Then $ |\chi_N| > \sum_{j=1}^{N-1} |\chi_j|$, so eq.\SLR \
cannot be satisfied.
	There are also countless examples that satisfy eq.\SLR.
	In the case of three spaces, for example, eq.\SLR \ states
that the largest genus is one less than the sum of the other two,
$g_3 = g_1 + g_2 -1$.
	We can get examples by taking arbitrary positive integers
for $g_1$ and $g_2$.
	The existence of solutions to eq.\SLR \ does not readily mean
that the corresponding amplitudes are non-zero.
	However, we will see in the next subsection
that eq.\SLR \ does not leave much room for improvement,
as long as we ask only the spatial topologies but no further details
of the boundary states.

\section{Transition paths}

	We shall present examples of transition paths for topology changes.
	Here to give such a path means to give the pair $(M,\omega)$
of a space-time manifold $M$ and a flat connection $\omega$
in an SO(2,1) bundle $P$ over $M$ satisfying the space condition \SLC \
for every boundary component $\Sigma_j$ of $M$.
	The amplitude $\IM$ is non-vanishing
for $\omega_j =\omega |_{\Sigma_j}$, and is within the spatial sector.

	We first define 3-manifold $M_k$, with $k$ an integer $\geq 2$.
	The aim is to give paths connecting $k+1$ spaces
$\Sigma_1, \ldots ,\Sigma_{k+1}$
consisting of $g=2$ surfaces $\Sigma_1, \ldots , \Sigma_k$ and
a $g=k+1$ surface $\Sigma_{k+1}$.
	This topology combination satisfies the selection rule \SLR:
$\chi_{k+1} = -2k = \sum_{j=1}^k \chi_j$.
	We represent $M_k$ as $M_k = \Lambda_k \cup V^k$.
	Fig.2 depicts $\Lambda_k$ and $V^k$ separately.
	$\Lambda_k$ is obtained from a $g=2k$ handlebody
(bounded by $\Gamma_k$)
by removing $g=2$ handlebodies (bounded by $\Sigma_1, \ldots ,\Sigma_k$),
while $V^k$ from a reflected copy of the $g=2k$ handlebody (bounded by $L^k$)
by removing a $g=k+1$ handlebody (bounded by $\Sigma_{k+1}$).
	In this, the two-handled surfaces $\Sigma_j \; (1 \leq j \leq k)$
pair the holes of $\Gamma_k$ (numbered from 1 to $2k$ in fig.2)
each handle linking one hole like $(1|2)(3|4) \cdots (2k-1|2k)$,
while the relation of the handles of $\Sigma_{k+1}$ to the holes of $L^k$
is like $(1|2,3|4,5| \cdots | 2k-2,2k-1|2k)$.
	The outer surfaces $\Gamma_k$ and $L^k$,
one being a reflected copy of the other, are identified
so that they represent a single surface in the interior of $M$.

	So we have manifold $M_k$.
	Before giving an example of flat connection $\omega$,
we show that $\IM$ is not divergent in the spatial sector.
	For simplicity we will work on the $k=2$ case.
	See fig.3.
	The fundamental group of $M_2$ can be described
by the generators $\a_1, \ldots, \a_8 \in \pi_1(M_2,*)$,
represented by the paths numbered in fig.3,
and the relations,
\global\advance\equanumber by 1 		
\xdef\RELa{(\chapterlabel \number\equanumber a)}
\xdef\RELb{(\chapterlabel \number\equanumber b)} 
\xdef\REL{(\chapterlabel \number\equanumber)} 	
$$
[\a_1,\a_2][\a_3,\a_4] = [\a_5,\a_6][\a_7,\a_8] = 1,	\eqno\RELa
$$
$$
		\a_5   = \a_4\a_3\a_4^{-1}, 		\eqno\RELb
$$
where the commutator $[\;,\;]$ is defined by $[\a,\b] = \a \b \a^{-1} \b^{-1}$.
	Let $\r: \pi_1(M,*) \rightarrow {\rm SO(2,1)}$ be
a holonomy homomorphism of a flat connection $\omega$ on $M$,
and let $\r_j$ denote the image of $\a_j$ under $\r$:
$\r_j = \r (\a_j) \in {\rm SO(2,1)}$.
	The gauge equivalence class of the ordered set $(\r_1,\r_2,\r_3,\r_4)$
is determined by $\omega_1$ the restriction of $\omega$ to $\Sigma_1$.
	Similarly, ${\rm cls}(\r_5, \r_6, \r_7, \r_8)$
is determined by $\omega_2$,
and ${\rm cls}(\r_1, \r_2, \r_3, \r_6\r_4, \r_7, \r_8)$ by $\omega_3$.
	If $\Sigma_1$ with $\omega_1$ corresponds to a state
in the spatial sector, then its holonomy homomorphism gives
a discrete embedding of $\pi_1(\Sigma) $ into
$ {\rm PSL(2,\R)} \approx {\rm SO(2,1)}$. From
 this it can be shown that if
$\g \in {\rm PSL(2,\R)}$ satisfies
$ \g\r_1\g^{-1} = \r_1 $ and $\g\r_2\g^{-1} = \r_2$,
then $\g=1$.
	Hence with the help of $\RELb$,
$\omega_1$ and $\omega_3$ together determine ${\rm cls}(\r_j| 1\leq j\leq 8)$
and therefore ${\rm cls}\omega$, completely.
	In particular, for given $\omega_1, \omega_2, \omega_3$
in the spatial sector, {\rm cls} $\omega$ is unique if it exists.
	Hence $\IM$ is divergence-free in the spatial sector.

	We now give an example of $\omega$, again concentrating on $k=2$.
	We leave $\omega_1$, or ${\rm cls}(\r_1, \ldots , \r_4)$,
arbitrary except that it belongs to the spatial sector.
	With a representative $(\r_1, \ldots ,\r_4)$,
we set
$$
\eqalign{
\r_5 &= \r_4\r_3\r_4^{-1}, \ \ \ \r_6 = \r_4, \cr
\r_7 &= \r_4\r_1\r_4^{-1}, \ \ \ \r_8 = \r_4\r_2\r_4^{-1}. \cr
}\eqn\DOO
$$
	Then $\r_j$, $1 \leq j \leq 8$, satisfy all the relations \REL.
	We have to check that $\omega_2$ and $\omega_3$ are
in the spatial sector.
	It is easily done for $\omega_2$, since
${\rm cls} (\r_1,\ldots,\r_4) = {\rm cls} (\r_5, \ldots, \r_8)$ by \DOO,
which means that $\Sigma_1$ and $\Sigma_2$ are `equivalent' under $\omega$.
	(The definition of the equivalence is as follows:
	$\omega_1$ is gauge equivalent to the pull-back of $\omega_2$
by some diffeomorphism of $\Sigma_1$ onto $\Sigma_2$
that preserves the orientation relative to that of $\partial M$.)
	To check on $\Sigma_3$, the easiest way would be to infer
${\rm eul}(P_j)$
for $j=3$ from the knowledge of the other two via eq.\VEC.
	Here we instead resort to the relation
of PSL(2,\R) flat connections on  $\Sigma$ of $g \geq 2$
to the Teichm\"uller space of $\Sigma$.
	A PSL(2,\R) connection on $\Sigma$ belongs
to an ${\rm eul} = \pm \chi_\Sigma$ bundle
if and only if its holonomy group $\subset {\rm PSL(2,\R)}$ is
the cover group of the Riemann surface $\Sigma$ with some complex structure.
	By eqs.\REL \ we have
$$
{\rm cls}
 (\r_1,\r_2,\r_3,\r_6\r_4,\r_7,\r_8)
={\rm cls}
 (\r_1,\r_2,\r_3,\r_4^2,\r_4\r_1\r_4^{-1},\r_4\r_2\r_4^{-1}).	\eqn\CLS
$$
	The left-hand side corresponds to the generators
of the holonomy group of $\omega_3$.
	Since $\Sigma_1$ with $\omega_1$ is spatial, $\r_1, \ldots, \r_4$
are the generators of the cover group of $\Sigma_1$ as a Riemann surface.
	Eq.\CLS \ implies that the holonomy group of $\omega_3$
is the cover group of a $g=3$ Riemann surface,
namely a double cover of $\Sigma_1$ (the one associated with a cut along
a path in the path class $\a_3$ if the basepoint $*$ is chosen on $\Sigma_1$).
	Hence $\Sigma_3$ is also in the spatial sector.
	We also see that $\Sigma_3$ is
`equivalent' to a $(-2)$-fold cover of $\Sigma_1$ under $\omega$,
by which we mean $\omega_3$ is gauge equivalent to the pull-back of $\omega_2$
by a double-cover map of $\Sigma_3$ onto $\Sigma_1$
that {\sl reverses} the orientation relative to that of $\partial M$.

	In the same way, we can define examples of flat connections on $M_k$
for general $k \geq 2$ that satisfy the spatial condition for the boundary.
	In these examples,
$\Sigma_1$ can be set in an arbitrary spatial state by selecting $\omega$,
but under $\omega$ the $g = 2$ surface $\Sigma_j$, $1 \leq j \leq k$,
are all equivalent
and the $g = k+1 $ surface $\Sigma_{k+1}$ is equivalent to a $(-k)$-fold
cover of $\Sigma_1$.

	We step further and put together the above transition paths
to seek more variety in the combinations of spatial topologies.
	Take an arbitrary set of spaces $\Sigma^\l \; (1 \leq \ell \leq N)$
of genus $g_\ell \geq 2$ that satisfy the selection rule \SLR.
	We construct a transition path for these spaces.
	We define $K\l = g_\ell -1$, so that eq.\SLC \ reads
$$
\sum\nolimits_{\ell=1}^N \epsilon_\ell K\l = 0.			\eqn\COK
$$
	For each $\ell$ with $K\l \geq 2 $,
 we take a copy of $M_{K\l}$, denote it by $M^\l$,
and identify $\Sigma^\l$ with the $g=K\l +1 $ boundary component
of $M^\l$: $ \Sigma^\l = \Sigma^\l_{K\l + 1} $.
	A set of copies of the previous transition paths
$(M^\l, \omega^\l), K\l \geq 2$,
can be chosen so that they have diffeomorphisms $\varphi_j^\l$ from
a standard $g=2$ surface $\Sigma_\infty$
onto $\Sigma_j^\l, 1 \leq j \leq K\l$,
which have the following properties.

\noindent
(i) With respect to a fiducial orientation of $\Sigma_\infty$,
the orientation induced from that of $\partial M^\l$
via $\varphi_j^\l: \Sigma_\infty \rightarrow \Sigma_j^\l$
is positive or negative according to whether $\epsilon_\ell = +1$ or $ -1$.
\hfil \break
\noindent
(ii) The flat connections $\varphi_j^{\l \ast} \omega^\l$ on $\Sigma_\infty$
are mutually gauge equivalent for all possible
$j$ and $\ell \; (1 \leq j \leq K\l, K\l \geq 2) $.

\hfil \break
\noindent
	Then for any $\ell, m$ with $\epsilon_\ell \not= \epsilon_m$,
we can join $( M^\l, \omega^\l)$ and $( M^{(m)}, \omega^{(m)})$
by pasting together a pair of $g=2$ boundary components
$\Sigma_i^\l$ and $\Sigma_j^{(m)}$ say,
by the identification map $\varphi_j^{(m)} \circ \varphi_i^{\l-1}$
and identifying $\omega^\l$ and $\omega^{(m)}$
across the junction by appropriate transition functions.
	Eq.\COK \ ensures that we can construct
a transition path $(M, \omega)$ for the spaces
$\Sigma^{(1)}, \ldots ,\Sigma^{(N)}$ by first joining $(M^\l,\omega^\l)$
at some pairs of the $g=2$ boundary components with opposite $\epsilon_\ell$
as above,
and then identifying  $\Sigma^\l$ of $K\l = 1$ with
the remaining $g=2$ components.

	Thus we came by a fairly systematic way
to construct a family of transition paths
which cover all the combinations of spatial topologies
that satisfy eq.\SLR \ and $\chi_j \leq -2 \; (g_j \geq 2)$.
	In particular we proved
that for $g_j \geq 2$ surfaces the selection rule \SLR \
is sufficient for $g_j \geq 2$ surfaces to admit an amplitude $\IM$
not identically zero in the spatial sector.
	We do not claim that the $\IM$ is finite.
	However, in the $\IP$ with $P$ arising in our examples,
infrared divergences can occur only through intermediate states
in the spatial sector.
	This is no more than a fancy way of saying that $P$ breaks down
into $P^\l$ over $M^\l \; (1 \leq \ell \leq N)$
with which each ${\cal I}_{P^\l}$ is free of infrared divergences.
	Finally we note that the existence of transition paths
can alternatively be proved by Horowitz' pull-back construction\refmark\SEEHOR.
	The topology of the spaces and flat connections on them
expected from this construction contain those we obtained above.
	This is remarkable, but we do not know
the precise relation of our examples
to the theoretical construction of Horowitz'.

\chapter{Discussions}

We start with comparing the selection rule \SLR \ with that of
Sorkin's\refmark\SEESOR.
The latter states that in a Lorentzian cobordism
the Euler characteristics of the initial and final spaces
are the same:
$$
\chi_{\rm in}=\chi_{\rm out}.					 \eqn\SSR
$$
A Lorentzian cobordism for spaces $S_{\rm in}, S_{\rm out}$
in Sorkin's definition consists of an interpolating
compact 3-manifold $M$ and a (non-degenerate) time-oriented
Lorentzian metric with respect to which $S_{\rm in}$ and $S_{\rm out}$
are respectively initial and final spatial boundaries.
The initial and final surfaces may have  more than one connected
component, but each component is required to be closed.
The metric is not required to solve the Einstein equation.
Lorentzian cobordisms may be regarded as the paths to be
summed over in quantum gravity in the metric formulation.
Eq.\SSR\  is the condition for the transition amplitude with
$S_{\rm in}, S_{\rm out}$ to be non-zero in this formulation.
To compare it with our result, we restrict ourselves to the case
in which $M$ is orientable.
Obviously our rule \SLR \ is observed under \SSR.
We can understand this in the following way.
With the metric in a Lorentzian cobordism, we can construct a
bundle $P$ of SO(2,1) frames of tangents to M.
Then the proof for \SLR \ applies to $P.$
The assumption that $P$ satisfies the space condition for the boundary
is justified by a slightly modified version
of the argument in sect.2.3.2.
Furthermore a little inspection of the same argument shows
 that whether a component
belongs to the initial side or to the final one
is decided by the sign
of ${\rm eul}(P_j)$ that appears in eq.\VEC.
Thus  \SLR \ is valid here and Sorkin's result can be reproduced
when the boundary components are $g\ge 2$ surfaces.

What we have found in sect.3 is that the selection rule from
Lorentzian cobordism still holds in our formulation save for the distinction
between in-coming and out-going states.
Actually we did not make an issue of the way of dividing the external
states into in- and out-states.
Since our quantum space-time does not admit a Lorentzian metric,
we cannot implement an in-out distinction
the way Sorkin does.
Of course we can define the distinction by
the sign of ${\rm eul}(P_j)$ with the orientation
induced from $\partial M.$
The justification is not easy to find, however.
There is no guarantee that
the orientation of a frame associated with $P$ relates to the orientation
of the base manifold $M,$ in contrast to the case of
Lorentzian cobordism where frames can be regarded as tangent to $M$
everywhere.
A plausible alternative is to use the asymmetry of
the asymptotic space-time
in the temporal direction.
A space-time with the topology $\Sigma\times \R$ either
has an initial singularity
and is future complete  or has
the same features reversed in time\refmark\SEEMES.
The asymptotic space-times complete in the direction away from
the interacting region
may be called out-going, and the others in-coming.
This way of grouping will surely make sense in the classical limit.
We do not know if it is well-defined in quantum theory.
Even if it is, the direction of time in this sense may not be diagonal
in our representation.
In fact, the pull-back construction of Horowitz\refmark\SEEHOR\
can be used to show that SO(2,1) flat connections on the boundary do not
fix the direction of time for the asymptotic space-times.

Restrictions on the way of dividing the external states
into in- and out-groups, if they are justified, may mean a
great deal to topology change.
Just for the sake of illustration, let us assume a grouping
that would lead to eq.\SSR.
Then a
creation of $g\ge 2$ surfaces from
nothing or a combination of $g=1$ surfaces is forbidden.
A more subtle example is provided by the paths with $M_k$ in sect.3.3.
Take $M=M_2$, for simplicity.
In terms of the spatial topology, only two processes
are allowed under \SSR:
$\Sigma_1 \sqcup \Sigma_2$ $\longrightarrow$ $\Sigma_3,$ and
its reverse $\Sigma_3$ $\longrightarrow$ $\Sigma_1 \sqcup \Sigma_2.$
The amplitude for the first process almost always vanishes.
This is seen by noting that
the relation $(3.5b)$ requires $\rho_3$ and $\rho_5$ to be mutually
conjugate.
The states on $\Sigma_1$ and $\Sigma_2$ are related on the
support of $\IM$, and the subspace of such states has a non-zero
codimension in the space of all the possible initial states.
Thus the initial state must be prepared with an infinite degree of
precision for the transition to occur.
To put it the other way, the space  $\Sigma_1 \sqcup \Sigma_2$
is practically stable as far as the above process is concerned.
The same observation applies to the reversed process as well.
	On the other hand, if we remove the restriction in the grouping
of the external states, the transition path $(M_2, \omega)$ in sect.3.3
implies that a $g=2$ surface has a decay mode
$\Sigma_1 \longrightarrow \Sigma_2 \sqcup \Sigma_3$
( or $\Sigma_2 \longrightarrow \Sigma_1 \sqcup \Sigma_3$ )
with a finite probability for any state on the initial surface.
	Transition paths with $M_k$ are the paths for topology change
no matter what way we divide the external states.
	The point is that whether they represent topology changing paths
starting from reasonably generic states
depends on which external states are on the initial side.

The Gibbons-Hartle approach\refmark\SEEGH\ applied on 2+1 dimensions
as done by Fujiwara, Higuchi, Hosoya, Mishima, and Siino\refmark\SEEFUJ,  also
gives some results on topology change.
In a critical path for a tunneling, the junction of the Riemannian and
Lorentzian regions can be made only across totally geodesic
surfaces.
Without a cosmological constant, this means that $g\ge2$ surfaces
do not emerge from the interacting region.
This may have something to do with the restriction we found on such
spacelike surfaces. The agreement is not so sharp since
we did find non-vanishing amplitudes, but we have to allow for
the fact that the tunneling approach is only an approximation.
In the case of a negative cosmological constant,
critical paths are possible only with $g\ge 2$ surfaces
but a large number of examples have been found\refmark\SEEFUJ.
Some of the examples violate the selection rule \SLR.
This may be reflecting the fact that the approach has its
foundation in Euclidean theory rather than Lorentzian.
The discrepancy may not be so great however, since
the critical paths are possible only for a countable number
of points in the moduli space for hyperbolic surfaces.

\beginsection
Acknowledgments

We would like to thank P.~Crehan, A.~Hosoya, N.~Sakai, and M.~Siino for
helpful comments.
We are also grateful to Y.~Fujiwara for bringing  ref.\refmark\SEEHOR \
to our attention.

\vfil
\eject
\refout		
\vfil
\eject

\centerline{{\fourteenrm FIGURE CAPTIONS}}

{\bf Figure~1} \
A 3-manifold $M$ for which $\IM$ is zero in the spatial sector.
It is represented as a handlebody of genus $g=2$
with two $g=1$ handlebodies removed.
Double-torus $\Sigma_3$ is on the outer boundary of $M$,
while the tori $\Sigma_1$ and $\Sigma_2$ inside $\Sigma_3$ constitute
the inner boundary.
The amplitude vanishes essentially because
in $M$ the loop $\gamma$ on $\Sigma_3$ shrinks to a point.

{\bf Figure~2} \
The interpolating manifold $M_k$
with a non-zero transition amplitude within the spatial sector.
$M_k$ is obtained by gluing $V^k$ and $\Lambda_k$ together
around boundary components $L^k$ and $\Gamma_k$.

{\bf Figure~3} \
The manifold $M_2$ drawn in the same way as in fig.~2.
The closed paths which represent the generators of $\pi_1(M_2,*)$
are also shown.

{\bf Figure~4} \
A picture of an interpolating manifold for spaces
$ \Sigma_\l$, $1 \le \ell \le 5$,
with
$ K(1)=K(3)=4, K(2)=K(4)=1, K(5)=2$.
We take a copy of $M_2$ and two copies of $M_4$
and then sewing together the manifolds and the connections on them
appropriately across $g=2$ surfaces.
The sign of $\epsilon_{\ell}$ for $\ell =1,3,5$ is indicated
in the centre of corresponding $M_{\ell}.$

\vfill

\end